
\overfullrule=0pt

\magnification=1200

\def\CSF{{\sl Chaos, Solitons and Fractals}}

\def\IEEETIT{{\sl IEEE Trans.\ Inform.\ Theory}}

\def\Na{{\sl Nature\/}}
\def\Naw{{\sl Naturwissenschaften}}

\def\PRA{{\sl Phys.\ Rev.\ A\/}}

\def\PRL{{\sl Phys.\ Rev.\ Lett.}}
\def\PRSLA{{\sl Proc.\ Roy.\ Soc.\ Lond.\ A\/}}

\def\PTRSLA{{\sl Phil.\ Trans.\ Roy.\ Soc.\ Lond.\ A\/}}

\def\Sc{{\sl Science\/}}
\def\SIAMJC{{\sl SIAM J. Comput.}}

\def\SIAMR{{\sl SIAM Rev.}}

\def\dajm{\hbox{D. A. Meyer}}

\def\schrodinger{\hbox{E. Schr\"odinger}}

\def\hfb{\hfil\break}

\catcode`@=11
\newskip\ttglue

   \font\ninerm=cmr9    \font\eightrm=cmr8   \font\sixrm=cmr6
  \font\ninebf=cmbx9   \font\eightbf=cmbx8  \font\sixbf=cmbx6
  \font\nineit=cmti9   \font\eightit=cmti8  
  \font\ninesl=cmsl9   \font\eightsl=cmsl8  
  \font\ninemi=cmmi9   \font\eightmi=cmmi8  \font\sixmi=cmmi6

\font\bigten=cmr10 scaled\magstep2 

\def\ninepoint{\def\rm{\fam0\ninerm}%
  \textfont0=\ninerm \scriptfont0=\sixrm
  \textfont1=\ninemi \scriptfont1=\sixmi
  \textfont\itfam=\nineit  \def\it{\fam\itfam\nineit}%
  \textfont\slfam=\ninesl  \def\sl{\fam\slfam\ninesl}%
  \textfont\bffam=\ninebf  \scriptfont\bffam=\sixbf
    \def\bf{\fam\bffam\ninebf}%
  \tt \ttglue=.5em plus.25em minus.15em
  \normalbaselineskip=11pt
  \setbox\strutbox=\hbox{\vrule height8pt depth3pt width0pt}%
  \normalbaselines\rm}

\def\eightpoint{\def\rm{\fam0\eightrm}%
  \textfont0=\eightrm \scriptfont0=\sixrm
  \textfont1=\eightmi \scriptfont1=\sixmi
  \textfont\itfam=\eightit  \def\it{\fam\itfam\eightit}%
  \textfont\slfam=\eightsl  \def\sl{\fam\slfam\eightsl}%
  \textfont\bffam=\eightbf  \scriptfont\bffam=\sixbf
    \def\bf{\fam\bffam\eightbf}%
  \tt \ttglue=.5em plus.25em minus.15em
  \normalbaselineskip=9pt
  \setbox\strutbox=\hbox{\vrule height7pt depth2pt width0pt}%
  \normalbaselines\rm}

\def\sfootnote#1{\edef\@sf{\spacefactor\the\spacefactor}#1\@sf
      \insert\footins\bgroup\eightpoint
      \interlinepenalty100 \let\par=\endgraf
        \leftskip=0pt \rightskip=0pt
        \splittopskip=10pt plus 1pt minus 1pt \floatingpenalty=20000
        \parskip=0pt\smallskip\item{#1}\bgroup\strut\aftergroup\@foot\let\next}
\skip\footins=12pt plus 2pt minus 2pt
\dimen\footins=30pc

\def\ie{{\it i.e.}}

\def\and{{\eightpoint AND}}

\def\DeutschJozsa{1}
\def\BernsteinVazirani{2}
\def\Simon{3}
\def\Grover{4}
\def\TerhalSmolin{5}
\def\Einstein{6}
\def\JozsaEkert{7}
\def\NMRexp{8}
\def\separability{9}
\def\Lloyd{10}
\def\AWB{11}
\def\Schumacher{12}
\def\Dirac{13}
\def\CEMM{14}
\def\BrassardHoyer{15}
\def\Schrodinger{16}
\def\TielkingJones{17}
\def\Rydberg{18}
\def\Freedman{19}
\def\QECC{20}
\def\faulttolerance{21}
\def\Shor{22}
\def\Knight{23}
\def\websearch{24}
\def\infovector{25}
\def\Grovertwo{26}
\def\BBBV{27}

\input epsf.tex

\dimen0=\hsize \divide\dimen0 by 13 \dimendef\chasm=0
\dimen1=\hsize \advance\dimen1 by -\chasm \dimendef\usewidth=1
\dimen2=\usewidth \divide\dimen2 by 2 \dimendef\halfwidth=2
\dimen3=\usewidth \divide\dimen3 by 3 \dimendef\thirdwidth=3
\dimen4=\hsize \advance\dimen4 by -\halfwidth \dimendef\secondstart=4
\dimen5=\halfwidth \advance\dimen5 by -10pt \dimendef\indenthalfwidth=5
\dimen6=\thirdwidth \multiply\dimen6 by 2 \dimendef\twothirdswidth=6
\dimen7=\twothirdswidth \divide\dimen7 by 4 \dimendef\qttw=7
\dimen8=\qttw \divide\dimen8 by 4 \dimendef\qqttw=8
\dimen9=\qqttw \divide\dimen9 by 4 \dimendef\qqqttw=9
\dimen10=2.25truein \dimendef\sciencecol=10

\parskip=0pt\parindent=0pt

\line{\hfil 25 January 2000}
\line{\hfil {\it revised\/} 22 June 2000}
\line{\hfill quant-ph/0007070}
\vfill
\centerline{\bf\bigten SOPHISTICATED QUANTUM SEARCH}
\medskip
\centerline{\bf\bigten WITHOUT ENTANGLEMENT}
\bigskip\bigskip
\centerline{\bf David A. Meyer}
\bigskip 
\centerline{\sl Project in Geometry and Physics}
\centerline{\sl Department of Mathematics}
\centerline{\sl University of California/San Diego}
\centerline{\sl La Jolla, CA 92093-0112}
\centerline{dmeyer@chonji.ucsd.edu}
\smallskip
\centerline{\sl and Institute for Physical Sciences}
\centerline{\sl Los Alamos, NM}
\smallskip

\vfill
\centerline{ABSTRACT}
\bigskip
\noindent Although entanglement is widely considered to be necessary 
for quantum algorithms to improve on classical ones, Lloyd has 
observed recently that Grover's quantum search algorithm can be 
implemented without entanglement, by replacing multiple particles with 
a single particle having exponentially many states.  We explain that 
this maneuver removes entanglement from {\sl any\/} quantum 
algorithm.  But {\sl all\/} physical resources must be accounted for 
to quantify algorithm complexity, and this scheme typically incurs 
exponential costs in some other resource(s).  In particular, we 
demonstrate that a recent experimental realization requires 
exponentially increasing precision.  There is, however, a quantum 
algorithm which searches a `sophisticated' database (not unlike a Web 
search engine) with a single query, but which we show does not require 
entanglement even for multiparticle implementations.
\bigskip\bigskip
\noindent 1999 Physics and Astronomy Classification Scheme:
                   03.67.Lx, 
                   32.80.Rm. 

\noindent 2000 American Mathematical Society Subject Classification:
                   81P68,    
                   68Q15.    

\smallskip
\global\setbox1=\hbox{Key Words:\enspace}
\parindent=\wd1
\item{Key Words:}  quantum algorithms, entanglement, query complexity.

\vfill
\eject

\headline{\ninepoint\it Sophisticated quantum search without 
                        entanglement            \hfill David A. Meyer}

\parskip=10pt
\parindent=20pt

Quantum algorithms must exploit some physical resource unavailable to
classical computers in order to solve problems in fewer steps
[\DeutschJozsa,\BernsteinVazirani,\Simon,\Grover,\TerhalSmolin].  
Entanglement, which seems the ``spookiest'' [\Einstein] to many 
people, has been argued to be the crucial quantum mechanical resource 
[\JozsaEkert].  This belief informs, for example, the criticism 
that NMR experiments performed to date [\NMRexp] have not actually 
realized quantum algorithms because at each timestep the state of the 
system can be described as a probabilistic ensemble of unentangled 
quantum states [\separability].  Lloyd [\Lloyd] and Ahn, Weinacht \& 
Bucksbaum [\AWB] have recently suggested, however, that entanglement 
is {\sl not\/} necessary for Grover's quantum search algorithm 
[\Grover].  In this Letter we clarify the situation by demonstrating 
that, contrary to their claims, the experimental realization of Ahn, 
Weinacht \& Bucksbaum [\AWB] requires an exponentially increasing 
amount of a resource---precision---replacing entanglement.  But we do 
not conclude from this that entanglement (or some replacement 
resource) is required.  Rather, we make the new and surprising 
observation that efficient quantum search of a `sophisticated' 
database (not unlike a Web search engine) requires no entanglement at 
any timestep:  a quantum-over-classical reduction in the number of 
queries is achieved using only interference, not entanglement, within 
the usual model of quantum computation.

The problem which forms the context for our discussion is database
search---identifying a specific record in a large database.  Formally,
label the records $\{0,1,\ldots,N-1\}$, where, for convenience when we
write the numbers in binary, we take $N = 2^n$ for $n$ a positive 
integer.  Grover considered databases which when queried about a 
specific number, respond only that the guess is correct or not 
[\Grover].  On a classical reversible computer we can 
implement a query by a pair of registers $(x,b)$ where $x$ is an 
$n$-bit string representing the guess, and $b$ is a single bit which 
the database will use to respond to the query.  If the guess is 
correct, the database responds by adding 1 (mod 2) to $b$; if it is 
incorrect, it adds 0 to $b$.  That is, the response of the database is 
the operation:  $(x,b) \to \bigl(x,b\oplus f_a(x)\bigr)$, where 
$\oplus$ denotes addition mod 2 and $f_a(x) = 1$ when $x = a$ and 0 
otherwise.  Thus if $b$ changes, we know that the quess is correct.  
Classically, it takes $N-1$ queries to solve this problem with 
probability 1.

Quantum algorithms work by supposing they will be realized in a 
quantum system, such as those described by Lloyd [\Lloyd] and Ahn,
Weinacht \& Bucksbaum [\AWB], which can be in a superposition of 
`classical' states.  These states form a basis for the Hilbert space
whose elements represent states of the quantum system.
The simplest such system is a {\sl qubit\/} [\Schumacher], which can 
be in a superposition of the states of a classical bit, \ie, 0 and 1.  
More generally, Grover's algorithm works with quantum queries which 
are linear combinations $\sum c_{x,b} |x,b\rangle$, where the 
$c_{x,b}$ are complex numbers satisfying $\sum |c_{x,b}|^2 = 1$ and 
$|x,b\rangle$ is Dirac notation [\Dirac] for the quantum state which 
represents the classical state $(x,b)$ of the two registers.  
The operations in quantum algorithms are unitary transformations, the 
quantum mechanical generalization of reversible classical operations.  
Thus the operation of the database Grover considered is implemented on 
superpositions of queries by a unitary transformation 
($f_a$-controlled-{\eightpoint NOT}), which takes $|x,b\rangle$ to 
$|x,b\oplus f_a(x)\rangle$.  Fig.~1 illustrates a quantum circuit 
implementation [\CEMM] of Grover's algorithm [\Grover].  Using 
$\lfloor{\pi\over4}\sqrt{N}\rfloor$ quantum queries it identifies the 
answer with probability close to 1:  the final vectors for the $N$ 
possible answers $a$ are nearly orthogonal.

\moveright\secondstart\vtop to 0pt{\hsize=\halfwidth
\vskip -0.5\baselineskip
$$
\epsfxsize=\sciencecol\epsfbox{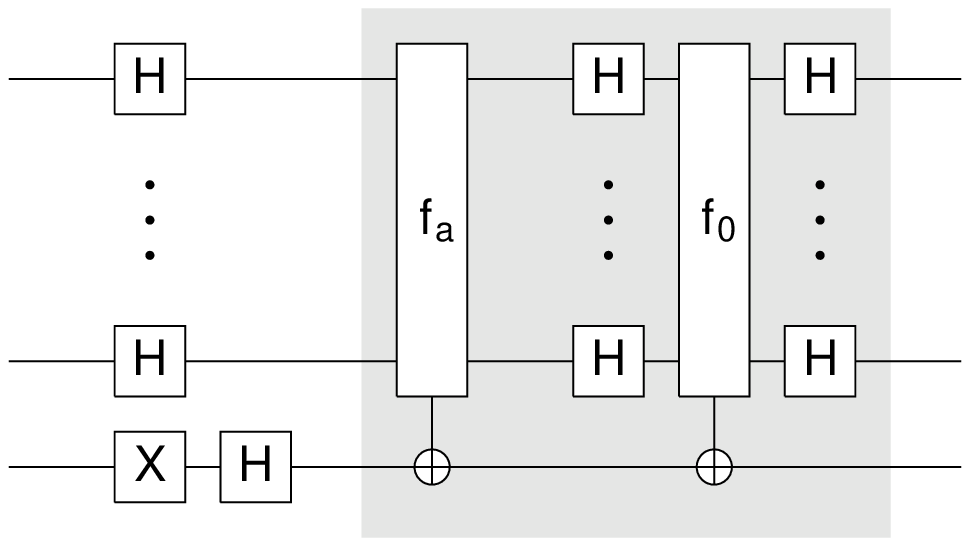}
$$
\vskip 0\baselineskip
\eightpoint{%
\noindent{\bf Figure~1}.  A (schematic) quantum circuit implementing 
Grover's algorithm.  Each horizontal line represents a single qubit, 
which is initialized (on the left) in state $|0\rangle$.  The portion 
of the circuit enclosed in the grey square is repeated 
$\lfloor{\pi\over4}\sqrt{N}\rfloor$ times and then the top $n$ qubits 
are measured.  $H$ is the Hadamard transformation 
$\bigl({1 \atop 1} {1 \atop -1}\bigr)/\sqrt{2}$, $X$ is the 
Pauli matrix $\bigl({0 \atop 1} {1 \atop 0}\bigr)$, and the `gates' 
acting on all $n+1$ qubits are $f_a$- and 
$f_0$-controlled-{\sixrm NOT} transformations, respectively.  As was
first noted by Brassard \& H{\o}yer, and subsequently by Grover, the 
$H^{\otimes n}$ ($\otimes I_2$) transformation conjugating the 
$f_0$-controlled-{\sixrm NOT} gate can be replaced by almost any 
unitary transformation [\BrassardHoyer].  Our discussion is 
independent of this choice.
}}
\vskip -\baselineskip
\parshape=22
0pt \halfwidth
0pt \halfwidth
0pt \halfwidth
0pt \halfwidth
0pt \halfwidth
0pt \halfwidth
0pt \halfwidth
0pt \halfwidth
0pt \halfwidth
0pt \halfwidth
0pt \halfwidth
0pt \halfwidth
0pt \halfwidth
0pt \halfwidth
0pt \halfwidth
0pt \halfwidth
0pt \halfwidth
0pt \halfwidth
0pt \halfwidth
0pt \halfwidth
0pt \halfwidth
0pt \hsize
This quantum circuit acts on an initial state 
$\psi_0 = |0\rangle\cdots|0\rangle|0\rangle 
        = |0\ldots0,0\rangle$.  The first set of gates transforms the
state to
$\psi_1 = {1\over\sqrt{2}}(|0\rangle + |1\rangle)\otimes\cdots\otimes
          {1\over\sqrt{2}}(|0\rangle + |1\rangle)\otimes\break
          {1\over\sqrt{2}}(|0\rangle - |1\rangle)
        = \sum_{x=0}^{N-1}|x\rangle
                          (|0\rangle - |1\rangle)/\sqrt{2N}$.  
Both these states, $\psi_0$ and $\psi_1$, are tensor products of the 
states of the individual qubits, so they are {\sl unentangled\/} 
[\Schrodinger].  This is no longer true for subsequent states of the 
system (except when $N=2$).  The last qubit, however, is never 
entangled with the others---after the first timestep it remains in 
state ${1\over\sqrt{2}}(|0\rangle - |1\rangle)$.
Lloyd's observation [\Lloyd], which is exploited by Ahn, Weinacht \& 
Bucksbaum [\AWB], is that the absence of entanglement in the $N=2$ 
case of Grover's algorithm (for which the guess register consists of a 
single qubit), generalizes to arbitrary $N$ if the guess register is 
realized by one $N$ state particle rather than by $n$ qubits.  In 
fact, Jozsa and Ekert [\JozsaEkert] made exactly 
this observation several years ago:  they wrote, ``The state of $n$ 
qubits is a $2^n$ dimensional space and can be isomorphically viewed 
as the state space of a {\sl single\/} particle with $2^n$ levels.  
Thus we simply view certain states of a single $2^n$ level particle as 
`entangled' via their correspondence under a chosen isomorphism 
between $\otimes^n {\cal H}_2$ and ${\cal H}_{2^n}$ (where 
${\cal H}_k$ denotes a Hilbert space of dimension $k$.)''.  So despite
the implication of Lloyd's [\Lloyd] and Ahn, Weinacht \& Bucksbaum's
[\AWB] papers, there is nothing special about Grover's algorithm:  
reformulating {\sl any\/} quantum algorithm this way, \ie, 
disregarding the tensor product structure of Hilbert space implicit in 
the use of qubits, removes entanglement from the system {\sl by 
definition}.  Nevertheless, one might hope that if a quantum 
algorithm---like Grover's---can be implemented naturally with a single 
particle, as Lloyd suggests [\Lloyd] and as Ahn, Weinacht \& Bucksbaum 
realized experimentally with $N$ Rydberg levels of a cesium atom 
[\AWB], there is some physical advantage to be gained.

But Jozsa and Ekert [\JozsaEkert] continue, 
``However the physical implementation of this correspondence appears 
always to involve an exponential overhead in some physical resource so 
that the isomorphism is {\sl not\/} a valid correspondence for 
considerations of complexity.'', again anticipating Lloyd's discussion 
[\Lloyd].  Although their data indicate that increasing $N$ requires 
more repetitions of the experiment to extract the answer [\AWB], 
Ahn, Weinacht \& Bucksbaum neglect the exponential overhead required 
for measurement and for realization of $N\times N$ unitary 
transformations:  They claim that extrapolation from their $N=8$ 
experiments to $N=20$ is straightforward and suggest that ultrafast 
shaped terahertz pulses [\TielkingJones] might realize more general 
unitary transformations than those used in their implementation of 
Grover's algorithm.  But because the difference (detuning) between 
adjacent Rydberg energy levels converges to 0 polynomially in $1/N$ 
(for $N$ labelling the energy levels) [\Dirac,\TielkingJones], 
both the laser pulses and the final measurements must be specified 
with exponentially increasing precision in $n$, the size of the 
problem [\Rydberg].  This should be contrasted with the standard model 
for quantum computation using {\sl poly-local\/} transformations 
implemented by polynomially many bounded size gates on Hilbert spaces 
with a tensor product decomposition [\Freedman];%
\sfootnote{$^*$}{The tensor factors need not be two dimensional, \ie, 
                 qubits.  Higher dimensional factors have been 
                 considered in the context of error correction 
                 [\QECC] and fault tolerance 
                 [\faulttolerance].  But in every case the 
                 dimension is bounded and scaling to larger problems
                 is achieved using polynomially many tensor factors.
                 Ahn, Weinacht \& Bucksbaum have demonstrated single
                 factor operations [\AWB]; gate operations analogous
                 to controlled-{\sixrm NOT} on two Rydberg atoms 
                 would be required for such an atomic system to 
                 realize quantum computation.}
these require 
specification of only polynomially many nontrivial amplitudes with 
constant precision.  As Bernstein \& Vazirani [\BernsteinVazirani] and 
Shor [\Shor] already emphasized in their original analyses of quantum 
models for computing, {\sl all\/} physical resources must be accounted 
for to quantify algorithm complexity; it is a mistake to ignore some 
because the requirements for them do not overwhelm small $N$ 
experiments.

Having identified an exponential cost associated with Ahn, Weinacht \& 
Bucksbaum's realization [\AWB] of Lloyd's suggestion for entanglement 
removal [\Lloyd], we are now ready to demonstrate that it is also a 
mistake to infer, as Lloyd's presentation might lead one to [\Knight], 
that quantum algorithms {\sl require\/} entanglement---or an 
exponential amount of some resource replacing it---to improve on 
classical algorithms.  Rather than Grover's `na{\"\i}ve' database, let 
us consider a `sophisticated' database which when queried about a 
specific number, responds with information about how close the guess 
is to the answer.  This kind of response is more like that returned 
by, for example, Web search engines, which typically order pages by 
relevance [\websearch].  A simple measure of relevance 
comes from the vector space model of information retrieval 
[\infovector]:  the records in the database and the guess are 
represented by vectors; then (the cosine of) the angle between a guess 
and any record measures their similarity and can be computed from the 
dot product of their vectors.  In our setting the `sophisticated' 
database acts on a query $(x,b)$ by computing the dot product of the 
$n$-dimensional binary vectors $x\cdot a$ and adding it to $b$ 
(mod 2).  Thus $(x,b) \to \bigl(x,b\oplus g_a(x)\bigr)$, where 
$g_a(x) = x\cdot a$.  Classically $n$ queries suffice to identify $a$ 
with probability 1.

Quantum mechanically, an underappreciated algorithm of Bernstein \& 
Vazirani [\BernsteinVazirani], rediscovered by Terhal \& Smolin 
[\TerhalSmolin], searches this `sophisticated' database with only 
{\sl a single\/} quantum query.%
\sfootnote{$^{\dagger}$}{It should be noted that the algorithm 
                 realized by 
                 Ahn, Weinacht \& Bucksbaum [\AWB] is not Grover's 
                 first [\Grover] which requires $O(\sqrt{N})$ queries,
                 but rather, something like his second [\Grovertwo] 
                 which implements
                 a single query on $O(N\log N)$ databases in parallel.
                 This number of databases is required to achieve 
                 sufficient statistical power to identify the 
                 solution, and is reflected in the increasing number
                 of times (commented on above) Ahn, Weinacht \& 
                 Bucksbaum need to repeat their experiment to extract
                 the answer as $N$ increases [\Rydberg].  This 
                 algorithm actually scales {\sl worse\/} than the 
                 classical algorithm.}
The operation of the database is 
implemented by the unitary transformation 
($g_a$-controlled-{\eightpoint NOT}) which takes $|x,b\rangle$ to 
$|x,b\oplus g_a(x)\rangle$.  A quantum circuit for their algorithm 
(slightly improved [\CEMM]) is shown in Fig.~2.  The first set of 
gates is the same as\break

\moveright\secondstart\vtop to 0pt{\hsize=\halfwidth
\vskip -3\baselineskip
$$
\epsfxsize=\sciencecol\epsfbox{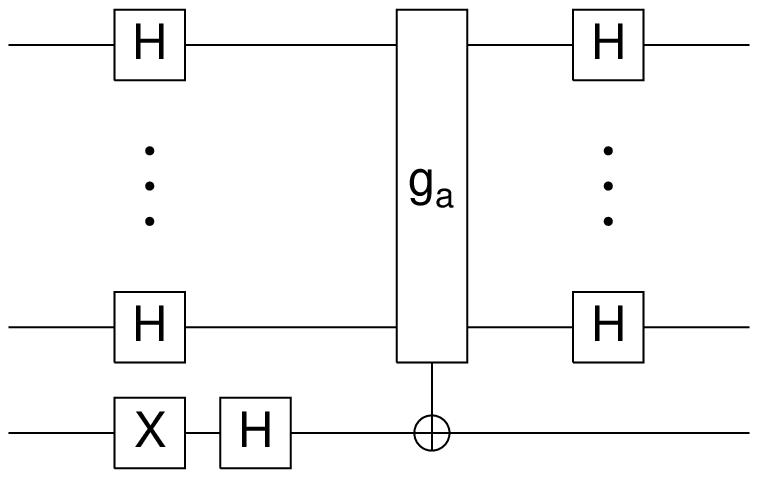}
$$
\vskip 0\baselineskip
\eightpoint{%
\noindent{\bf Figure~2}.  A (schematic) quantum circuit implementing 
Bernstein \& Vazirani's algorithm.  Each horizontal line again
represents a qubit, which is initialized (on the left) in state
$|0\rangle$.  $H$ and $X$ are as in Fig.~1 and the ``gate'' acting on 
all $n+1$ qubits is the $g_a$-controlled-{\sixrm NOT}
transformation of the `sophisticated' database.  The top $n$ qubits
are measured at the end of the circuit.
}}
\null\vskip -3.75\baselineskip
\parshape=16
0pt \halfwidth
0pt \halfwidth
0pt \halfwidth
0pt \halfwidth
0pt \halfwidth
0pt \halfwidth
0pt \halfwidth
0pt \halfwidth
0pt \halfwidth
0pt \halfwidth
0pt \halfwidth
0pt \halfwidth
0pt \halfwidth
0pt \halfwidth
0pt \halfwidth
0pt \hsize
\noindent in Fig.~1, and takes $\psi_0 = |0\ldots0,0\rangle$ to 
$\psi_1 = \sum_{x=0}^{N-1} |x\rangle
                           (|0\rangle - |1\rangle)/\sqrt{2N}$.  After 
the database responds to this quantum query, the state is 
$\psi_2 = \sum_{x=0}^{N-1} (-1)^{x\cdot a} |x\rangle
                           (|0\rangle - |1\rangle)/\sqrt{2N}$.
The final set of gates outputs 
$\psi_3 = |a\rangle\otimes{1\over\sqrt{2}}(|0\rangle - |1\rangle)$,
whereupon measuring the first ($n$-qubit) register identifies $a$ 
with probability 1 (the output states for different 
$a$\kern0.08333em{s} are orthogonal).  Comparing with Grover's 
algorithm, we recognize that the 
last qubit still remains unentangled with the first register, so that 
we could again implement the latter with a single $2^n$ state particle 
and have no entanglement at any timestep.  But this would be 
redundant:  {\sl there is no entanglement in Bernstein \& Vazirani's 
algorithm}.  To see this, observe that just as in Grover's algorithm
there is no entanglement in $\psi_0$ or $\psi_1$, and there is none in 
$\psi_3$, since $|a\rangle$ is simply a tensor product of qubits each 
in state $|0\rangle$ or $|1\rangle$.  But $\psi_3$ was obtained from 
$\psi_2$ by a unitary transformation acting on each of the $n+1$ 
qubits separately.  Such a unitary transformation cannot change the 
entanglement of a state, so $\psi_2$ must also be unentangled.  

To summarize:  any quantum algorithm in the usual poly-local model for 
quantum computing can be rewritten to have no entanglement at any 
timestep, simply by disregarding the tensor product structure of the 
Hilbert space.  Doing so physically incurs some exponential cost:  in 
energy, in measurement precision, or in specification of the required 
unitary transformations.  But one should not conclude that 
entanglement is {\sl required\/} for quantum-over-classical complexity 
reduction.  Without entanglement at any timestep, Bernstein \& 
Vazirani's quantum algorithm for `sophisticated' database search does 
not just reduce the number of queries required classically by a square 
root factor, but all the way from $n$ to 1.  Furthermore, we have 
shown for the first time that quantum interference alone suffices to 
reduce the query complexity of a problem within the standard model for 
quantum computation.  Since implementing the 
$g_a$-controlled-{\eightpoint NOT} `gate' with a subcircuit of 
{\sl local\/} gates would introduce entanglement at intermediate 
timesteps, however, one might conclude that counting queries (or, more 
generally, nonlocal function calls) is a poor way to study the power 
of quantum algorithms [\BBBV].  But it was Simon's algorithm [\Simon] 
(which exponentially reduces the number of nonlocal evaluations 
required to determine the period of a function) that led to Shor's 
quantum factoring algorithm [\Shor], so it seems more productive to 
understand quantum search of a `sophisticated' database as 
demonstrating the importance of interference and orthogonality, rather 
than entanglement, in quantum algorithms.  This perspective may 
contribute to discovering the new algorithms necessary for quantum 
computing to become more generally useful.

\medskip
\noindent{\bf Acknowledgements}
\nobreak

\nobreak
\noindent I thank Thad Brown, Mike Freedman, Raymond Laflamme, Melanie 
Quong, John Smolin and Nolan Wallach for useful discussions.  This 
work has been partially supported by Microsoft Research and the 
National Security Agency (NSA) and Advanced Research and Development 
Activity (ARDA) under Army Research Office (ARO) contract number 
DAAG55-98-1-0376.

\medskip
\global\setbox1=\hbox{[00]\enspace}
\parindent=\wd1

\noindent{\bf References}
\bigskip

\parskip=0pt
\item{[\DeutschJozsa]}
D. Deutsch,
``Quantum theory, the Church-Turing principle and the universal 
  quantum computer'',
\PRSLA\ {\bf 400} (1985) 97--117;\hfb
D. Deutsch and R. Jozsa,
``Rapid solution of problems by quantum computation'',
\PRSLA\ {\bf 439} (1992) 553--558.

\item{[\BernsteinVazirani]}
E. Bernstein and U. Vazirani,
``Quantum complexity theory'',
in 
{\sl Proceedings of the 25th Annual ACM Symposium on the Theory of
Computing}, San Diego, CA, 16--18 May 1993
(New York: ACM 1993) 11--20.
%

\item{[\Simon]}
D. R. Simon,
``On the power of quantum computation'',
in S. Goldwasser, ed.,
{\sl Proceedings of the 35th Symposium on Foundations of Computer 
Science}, Santa Fe, NM, 20--22 November 1994
(Los Alamitos, CA:  IEEE 1994) 116--123.
%

\item{[\Grover]}
L. K. Grover,
``A fast quantum mechanical algorithm for database search'',
in 
{\sl Proceedings of the 28th Annual ACM Symposium on the Theory of
Computing}, Philadelphia, PA, 22--24 May 1996
(New York:  ACM 1996) 212--219.

\item{[\TerhalSmolin]}
B. M. Terhal and J. A. Smolin,
``Single quantum querying of a database'',
\PRA\ {\bf 58} (1998) 1822--1826.

\item{[\Einstein]}
A. Einstein, 
in
{\sl The Born-Einstein Letters:  Correspondence between Albert
Einstein and Max and Hedwig Born from 1916 to 1955}, 
with commentaries by M. Born, transl.\ by I. Born
(New York:  Walker \& Co.\ 1971) p.\ 158.

\item{[\JozsaEkert]}
R. Jozsa,
``Entanglement and quantum computation'',
in
S. A. Huggett, L. J. Mason, K. P. Tod, S. T. Tsou and 
N. M. J. Woodhouse, eds.,
{\sl The Geometric Universe:  Science, Geometry, and the Work of 
Roger Penrose\/}
(Oxford:  Oxford University Press 1998) 369--379;\hfb
A. Ekert and R. Jozsa,
``Quantum algorithms:  entanglement-enhanced information 
  processing'',
\PTRSLA\ {\bf 356} (1998) 1769--1782.

\item{[\NMRexp]}
I. L. Chuang, L. M. K. Vandersypen, X. Zhou, D. W. Leung and S. Lloyd,
``Experimental realization of a quantum algorithm'',
\Na\ {\bf 393} (1998) 143--146;\hfb
D. G. Cory, W. Mass, M. Price, E. Knill, R. Laflamme, W. H. Zurek, 
T. F. Havel and S. S. Somaroo,
``Experimental quantum error correction'',
\PRL\ {\bf 81} (1998) 2152--2155;\hfb
J. A. Jones, M. Mosca and R. H. Hansen,
``Implementation of a quantum search algorithm on a nuclear
  magnetic resonance computer'',
\Na\ {\bf 393} (1998) 344--346.

\item{[\separability]}
S. L. Braunstein, C. M. Caves, R. Jozsa, N. Linden, S. Popescu and
R. Schack,
``Separability of very noisy mixed states and implications for NMR
  quantum computing'',
\PRL\ {\bf 83} (1999) 1054--1057;\hfb
%
%
For a response to this criticism, see
R. Laflamme,
``Review of `Separability of very noisy mixed states and 
  implications for NMR quantum computing'\thinspace'',
{\sl Quick Reviews in Quantum Computation and Information},
{\tt http://quickreviews.org/qc/}.

\item{[\Lloyd]}
S. Lloyd,
``Quantum search without entanglement'',
\PRA\ {\bf 61} (1999) 010301.

\item{[\AWB]}
J. Ahn, T. C. Weinacht and P. H. Bucksbaum,
``Information storage and retrieval through quantum phase'',
\Sc\ {\bf 287} (2000) 463--465.

\item{[\Schumacher]}
B. Schumacher,
``Quantum coding (information theory)'',
\PRA\ {\bf 51} (1995) 2738--2747.

\item{[\Dirac]}
P. A. M. Dirac,
{\sl The Principles of Quantum Mechanics}, fourth edition
(Oxford:  Oxford University Press 1958).

\item{[\CEMM]}
R. Cleve, A. Ekert, C. Macchiavello and M. Mosca,
``Quantum algorithms revisited'',
\PRSLA\ {\bf 454} (1998) 339--354.

\item{[\BrassardHoyer]}
G. Brassard and P. H{\o}yer,
``An exact quantum polynomial-time algorithm for Simon's problem'',
in
{\sl Proceedings of the Fifth Israeli Symposium on Theory of Computing
  and Systems},
Ramat-Gan, Israel, 17--19 June 1997
(Los Alamitos, CA:  IEEE 1997) 12--23;\hfb
L. K. Grover,
``Quantum computers can search rapidly by using almost any 
  transformation'',
\PRL\ {\bf 80} (1998) 4329--4332.

\item{[\Schrodinger]}
\schrodinger,
``{\it Die gegenw\"artige Situation in der Quantenmechanik\/}'',
\Naw\ {\bf 23} (1935) 807--812; 823--828; 844--849.

\item{[\TielkingJones]}
N. E. Tielking and R. R. Jones,
``Coherent population transfer among Rydberg states by 
  subpicosecond, half-cycle pulses'',
\PRA\ {\bf 52} (1995) 1371--1381.

\item{[\Rydberg]}
\dajm,
``Rydberg state manipulation is not quantum computation'',
UCSD preprint (2000).

\item{[\Freedman]}
M. H. Freedman,
``Poly-locality in quantum computing'',
{\tt quant-ph/0001077}.

\item{[\QECC]}
E. Knill, 
``Non-binary error bases and quantum codes'',
{\tt quant-ph/9608048};\hfb
E. M. Rains,
``Nonbinary quantum codes'',
\IEEETIT\ {\bf 45} (1999) 1827--1832

\item{[\faulttolerance]}
D. Aharonov and M. Ben-Or,
``Fault tolerant quantum computation with constant error'',
{\tt quant-ph/9611025};\hfb
D. Gottesman,
``Fault-tolerant quantum computation with higher-dimensional
  systems'',
\CSF\ {\bf 10} (1999) 1749--1758.

\item{[\Shor]}
P. W. Shor,
``Algorithms for quantum computation:  discrete logarithms and 
  factoring'',
in S. Goldwasser, ed.,
{\sl Proceedings of the 35th Symposium on Foundations of Computer 
Science}, Santa Fe, NM, 20--22 November 1994
(Los Alamitos, CA:  IEEE 1994) 124--134.
%

\item{[\Knight]}
P. Knight,
``Quantum information processing without entanglement'',
\Sc\ {\bf 287} (2000) 441--442.

\item{[\websearch]}
B. Yuwono and D. L. Lee,
``Search and ranking algorithms for locating resources on the World
  Wide Web'',
in
S. Y. W. Su, ed.,
{\sl Proceedings of the Twelfth International Conference on Data 
Engineering}, New Orleans, LA, 26 February--1 March 1996
(Los Alamitos, CA:  IEEE 1996) 164--171;\hfb
M. P. Courtois and M. W. Berry,
``Results ranking in Web search engines'',
{\sl Online\/} {\bf 23}:3 (1999) 39--46.

\item{[\infovector]}
G. Salton and M. McGill,
{\sl Introduction to Modern Information Retrieval\/}
(New York:  McGraw-Hill 1983);\hfb
M. W. Berry, Z. Drma\v c and E. R. Jessup,
``Matrices, vector spaces, and information retrieval'',
\SIAMR\ {\bf 41} (1999) 335--362.

\item{[\Grovertwo]}
L. K. Grover,
``Quantum computers can search arbitrarily large databases by a 
  single query'',
\PRL\ {\bf 79} (1997) 4709--4712.

\item{[\BBBV]}
C. H. Bennett, E. Bernstein, G. Brassard and U. Vazirani,
``Strengths and weaknesses of quantum computing'',
\SIAMJC\ {\bf 26} (1997) 1510--1523.

\bye